\documentclass[12pt,reqno]{article}
\usepackage{lipsum}
\usepackage[usenames]{color}
\usepackage{amssymb}
\usepackage{graphicx}
\usepackage{amscd}
\usepackage{blindtext}
\usepackage{marvosym}
\usepackage{tikz}
\usepackage{tikz}
\usepackage{tikz}
\usepackage{tikz-qtree}
\usepackage{mathtools}
\usepackage[most]{tcolorbox}

\tcbuselibrary{breakable}

\usepackage[colorlinks=true,
linkcolor=webgreen,
filecolor=webbrown,
citecolor=webgreen]{hyperref}

\definecolor{webgreen}{rgb}{0,.5,0}
\definecolor{webbrown}{rgb}{.6,0,0}

\usepackage{color}
\usepackage{fullpage}
\usepackage{float}

\usepackage{graphics,amsmath,amssymb}
\usepackage{amsthm}
\usepackage{amsfonts}
\usepackage{latexsym}
\usepackage{epsf}
\usepackage{MnSymbol}
\usepackage{tabularx}
\usepackage{amsthm}
\usepackage{enumitem}

\begin{document}

\bibliographystyle{plain}

\begin{center}
\epsfxsize=4in
\end{center}

\theoremstyle{plain}
\newtheorem{theorem}{Theorem}
\newtheorem{corollary}{Corollary}
\newtheorem{lemma}{Lemma}
\newtheorem{proposition}{Proposition}
\renewcommand{\theenumi}{\Alph{enumi}}

\theoremstyle{definition}
\newtheorem{definition}{Definition}
\newtheorem{example}{Example}
\newtheorem{remark}{Remark}

\newenvironment{claim}[1]{\par\noindent\underline{Claim:}\space#1}{}
\newenvironment{claimproof}[1]{\par\noindent\underline{Proof:}\space#1}{\hfill $\blacksquare$}

\begin{center}
\vskip 1cm{\LARGE\bf{Cryptanalysis of ITRU}}
\vskip 1cm
{\Large \bf  H.~R.~Hashim, A.~Molnár and Sz.~Tengely}
\vspace{3mm}

Institute of Mathematics, University of Debrecen\\
P. O. Box 400, 4002 Debrecen, Hungary\\
e-mail: \url{hashim.hayder.raheem@science.unideb.hu}\\
e-mail: \url{alexandra980312@freemail.hu}\\
e-mail: \url{tengely@science.unideb.hu}
\vspace{2mm}

\end{center}

\vskip .2 in

\begin{abstract}
ITRU cryptosystem is a public key cryptosystem and one of the known variants of NTRU cryptosystem.  Instead of working in a truncated polynomial ring, ITRU cryptosystem is based on the ring of integers. The authors claimed that ITRU has better features comparing to the classical NTRU, such as having a simple parameter selection algorithm, invertibility, and successful message decryption, and better security. In this paper, we present an attack technique against the ITRU cryptosystem, and it is mainly based on a simple frequency analysis on the letters of ciphertexts.
\end{abstract}

\section{Introduction}

The study of cryptography has been interested to cryptologists for long time because the necessity of transferring important information secretly, which established the existence of many types of cryptosystems. It is well-known that there are two types of cryptography, which are symmetric cryptography and asymmetric cryptography ( or, public key cryptography). In the symmetric cryptosystem, the same key being used in the encryption and decryption procedures. However, in the asymmetric cryptosystem two different keys are used; the public key that should be announced to everyone and the corresponding private key has to be secret. In fact, many models of these cryptosystems have been established by several cryptologists. Indeed, nowadays the most used cryptography is the public key cryptography for its better efficiency and security comparing to the other type. The security of many public key cryptosystems such as Rivest, Shamir and Adelman (RSA) cryptosystem \cite{MR700103}, McEliece cryptosystem \cite{1978DSNPR}, ElGamal cryptosystem \cite{MR798552}, or elliptic curve cryptosystem (ECC) \cite{MR866109} is based on different intractable mathematical problems. In practice, all of these public key cryptosystems are far slower than symmetric cryptosystems such as Data Encryption Standard (DES) cryptosystem \cite{standard1977l} or Advanced Encryption Standard (AES) cryptosystem \cite{advanced2001} in terms of space and computational complexity and for this reason they are often simply used to solve the problem of sharing a secret key for use in a symmetric cryptosystem (for more details, see \cite{MR2221458}, \cite{key:article}, and the references given there ). 

Therefore, the main target for cryptologists is the discovery of a fast public key cryptosystem based on different hard problems. In $1996,$ Hoffstein, Pipher and Silverman \cite{zbMATH01186948} proposed a class of fast public key cryptosystems called NTRU ($N^{\mbox{th}}$ Degree Truncated Polynomial Ring) cryptosystem, which was published in $1998.$ This cryptosystem is considered as a lattice-based public key cryptosystem, and it is the first asymmetric cryptosystem based on the polynomial ring $\frac{\mathbb{Z}[X]}{(X^N-1)}.$ Indeed, it has very good features comparing to other public key cryptosystems such  as reasonably short, easily created keys, high speed, and low memory requirements. Its encryption and decryption procedures rely on a mixing system presented by polynomial algebra combined with a clustering principle based on elementary probability theory. From its lattice-based structure, the security of the NTRU cryptosystem is based on the hardness of solving the Closest Vector Problem (CVP), which is a computational problem on lattices closely related to Shortest Vector Problem (SVP) and considered to be NP hard (non-deterministic polynomial-time hardness) (for more details, see \cite{Micciancio2005} and the references given there ).

In fact, the inventors  \cite{zbMATH01186948} proved that the NTRU cryptosystem preforms much faster than other public key cryptosystems. For instance, the encryption and decryption procedure of a message block of length $N$ takes $\mathcal{O}(N^2)$ operations using the NTRU cryptosystem and this is considerably faster than the $\mathcal{O}(N^3)$ operations required by RSA cryptosystem. Further, the key lengths of NTRU cryptosystem are $\mathcal{O}(N)$, which is very good comparing to the $\mathcal{O}(N^2)$ key lengths required by other fast public key cryptosystems presented in \cite{MR1630399} and \cite{1978DSNPR}. 

Furthermore, preliminary experimental results by Shen, Du, and Chen \cite{5341656} showed that the speed of the NTRU cryptosystem is much faster than that of the RSA cryptosystem in which the key generation is more than $200$ times faster, the encryption is almost $3$ times faster, and the decryption is about $30$ times faster. These results show the applicable possibility of NTRU cryptosystem in mobile Java systems.

For further enhancement of the security of the NTRU cryptosystem, researchers have been proposing several variants of NTRU cryptosystem. Starting with a generalization of NTRU cryptosystem proposed by Banks and Shparlinski \cite{zbMATH01955782} with non-invertible polynomials on the same ring as NTRU. The main advantage of this variant is that it is more secure against some of the known attacks on the original NTRU cryptosystem such as lattice attack. On the other hand, it is less efficient than NTRU since the lengths of its public key and the ciphertext are twice the ones in the classical NTRU cryptosystem.  Another analogue of NTRU cryptosystem was introduced by Gaborit, Ohler, and  Sol{\'e} \cite{gaborit:inria} called CTRU cryptosystem in which the ring $\mathbb{Z}$ in NTRU cryptosystem is replaced by the ring of polynomials $\mathbb{F}_2[T]$. A new variant of the NTRU cryptosystem was presented by Coglianese and Goi \cite{zbMATH05250294} called MaTRU cryptosystem. However, it operates under the same general principles as the NTRU cryptosystem, it works in a different ring with a different linear transformation in the encryption and decryption procedures. As a result, MaTRU cryptosystem is more efficient and has a better security level comparing to NTRU cryptosystem. Kouzmenko \cite{kouzmenko2006} used Gaussian integers instead of the ring $\mathbb{Z}$ in NTRU cryptosystem to propose a generalization of NTRU cryptosystem. However, it is not as efficient as NTRU, this scheme is slightly more secure against lattice attack than NTRU cryptosystem. By replacing the ring $\mathbb{Z}$ in NTRU cryptosystem by the Eisenstein integers $\mathbb{Z}[\zeta_3]$, Nevins, KarimianPour, and Miri \cite{zbMATH05723624} proposed another variant, which we they called it by ETRU cryptosystem, which presents a more difficult lattice problem for lattice attacks, for the same level of decryption failure security. Malekian, Zakerolhosseini, and Mashatan \cite{malekian2011qtru} presented a new variant called QTRU cryptosystem based on using the ring of quaternions instead of the ring $\mathbb{Z}$ in NTRU cryptosystem. They showed that the structure of QTRU cryptosystem gives more resistant to some lattice-based attacks comparing to the classical NTRU cryptosystem. 

Other variants have been introduced by many authors such ILTRU cryptosystem, which is a modification of ETRU cryptosystem, introduced by Karbasi and Atani \cite{eprint-2015-26958}. The security of this cryptosystem is based on the worst case hardness of the  approximate both SVP and CVP in ideal  lattices. 

Last but not least, we mention one of the known variants of NTRU cryptosystem called ITRU cryptosystem, which was presented in $2017$ by Gaithuru, Salleh, and Mohamad \cite{gaithuru2017itru}. Instead of working in a truncated polynomial ring, ITRU cryptosystem is based on the ring of integers. They showed that the ITRU cryptosystem has some interesting features such as having a simple parameter selection algorithm, invertibility, and successful message decryption. In fact, a description of a parameter selection algorithm and an implementation of ITRU with an application were provided. As a result, they claimed that the ITRU cryptosytem has a successful message decryption, which leads to more assurance of security in comparison to NTRU cryptosystem. Other variants of NTRU cryptosystem can be found, e.g. in \cite{10.1145}, \cite{articlebitru}, \cite{articlepairtru}, \cite{10.1007/978}, \cite{doi:10.1002/sec.1693}, \cite{camara2018dtru1}.

However, the inventors of NTRU cryptosystem ensured that it is  extremely unlikely to several potential attacks against the scheme to succeed ( particularly, the standard lattice-based attack ) since the secret key was surrounded by a“cloud” of exponentially many unrelated lattice vectors. Later, in $2001$ Coppersmith and Shamir \cite{10.1007/3-540-6} showed that the security of NTRU cryptosystem is not necessarily based on the difficulty of reducing the NTRU lattice since the lattice reduction can be one of the practical attacks against  NTRU cryptosystem. In fact, they presented a lattice-based attack, which can either find the original secret key $k$ or an alternative key $k'$ which can be used instead of $k$ to obtain the plaintexts by decrypting the corresponding ciphertexts with only slightly higher computational complexity. After that, many types of lattice-based attacks on the NTRU cryptosystem and its variants have been occurred. It is important to mention that all of these attacks have focused primarily on the “secret key recovery”  problem. For instance, Gentry \cite{zbMATH01722678} proposed lattice-based attacks that are especially effective when N, in the polynomial ring that used in the classical NTRU cryptosystem, is composite. He used low-dimensional lattices to find a folded version of the private key, where this key has $d$ coefficients where $d$ dividing $N.$ This folded private key is used to recover a folding of the plaintext, or it helps to recover the original private key.

However, a chosen ciphertext attack is another type of attacks, which was already used in \cite{zbMATH06590793} or \cite{joye1997importance} against other public key cryptosystems. Here, the attacker constructs invalid cipher messages. By knowing the plaintexts corresponding to his messages, she can get some information about the private key or even recover it. Such an attack was used against the NTRU cryptosystem by Jaulmes and Joux \cite{10.1007/3-54}. Similar attack to the later one was proposed by Meskanen and Renvalla \cite{zbMATH05021034}.

Another attack on NTRU cryptosystem hardware implementations, that employ scan based Design-for-Test (DFT) techniques, was proposed by Kamal and Youssef \cite{6329211}, and they called it a scan-based side channel attack. This attack determines the scan chain structure of the polynomial multiplication circuits used in the decryption algorithm which allows the cryptanalyst to efficiently retrieve the secret key.

More attack techniques against NTRU cryptosystem and its variants can be found, i.e. \cite{eprint-2003-11720}, \cite{zbMATH05242559}, \cite{zbMATH02091528}, \cite{10.1007/978-3-540-74143-5_9}, \cite{MR3989003}, and the references given there. 

In fact, most of the attacks against the NTRU cryptosystem especially the ones mentioned above focus primarily on the “secret key recovery”  problem. Therefore, in this paper we present a new attack technique to break the ITRU cryptosystem proposed in \cite{gaithuru2017itru}. Since the ITRU cryptosystem is a substitution cipher, so our attack is mainly based on a simple frequency analysis on the letters of ciphertexts using a function implemented in SageMath \cite{sage} as \texttt{frequency\_distribution()}. As a result, this techniques will recover the corresponding plaintexts immediately with no need of having the private keys.

\section{The ITRU Cryptosystem}

As mentioned earlier, instead of working in a truncated polynomial ring ITRU cryptosystem is based on the ring of integers. The parameters and the main steps of ITRU cryptosystem are as follows.
\begin{itemize}
  \item[$\square$] The value of $p$ is suggested to be 1000.
  \item[$\square$] Random integers $f, g$ and $r$ are chosen such that $f$ is invertible modulo $p$.
  \item[$\square$] A prime $q$ is fixed satisfying $q>p \cdot r \cdot g+f \cdot m$, where $m$ is the representation of the message in decimal form. The suggested conversion is based on $ASCII$ conversion tables, that is the one with $a \rightarrow 97$.
  \item[$\square$] One computes $F_p \equiv f^{-1} \pmod{p}$ and $F_q \equiv f^{-1} \pmod{q}$. These computations can be done by using the extended Euclidean algorithm.
  \item[$\square$] The public key is consisted of $h$ and $q$ such that 
\begin{equation}\label{Itru1}
h \equiv p \cdot F_q \cdot g \pmod{q}.
\end{equation}
 \item[$\square$] The encryption procedure is similar to the one applied in NTRU cryptosystem \cite{zbMATH01186948}, one generated a random integer $r$ and computes
\begin{equation}\label{Itru2}
e \equiv r \cdot h + m \pmod{q}.
\end{equation}
 \item[$\square$] To get the plaintext from the ciphertext one determines 
\begin{equation}\label{Itru3}
a \equiv f \cdot e \pmod{q}.
\end{equation}
\item[$\square$] Recovering the message is done by computing
\begin{equation}\label{Itru4}
F_p \cdot a \pmod{p}.
\end{equation}
\end{itemize}
In order to show this later recovery leads to the original plaintext at the end, one can show that as follows.
Combining equation \eqref{Itru3} with \eqref{Itru2} and \eqref{Itru1}, with use of  of the fact that $ f \cdot F_q \equiv 1 \pmod{q}$ we obtain that
\begin{equation}\label{Itru5}
a \equiv f \cdot e \equiv f \cdot ( r \cdot h + m) \equiv f \cdot ( r \cdot p \cdot F_q \cdot g +m) \equiv  r \cdot p \cdot g+  f \cdot m \pmod{q}.
\end{equation}
It remains to compute $F_p \cdot a \pmod{p}$ by substituting \eqref{Itru5} in \eqref{Itru4} and using the fact that  $ f \cdot F_p \equiv 1 \pmod{p}$. We obtain that 
$$
 F_p \cdot a \equiv F_p \cdot ( r \cdot p \cdot g+  f \cdot m) \equiv F_p \cdot f \cdot m  \equiv m \pmod{p}.
$$

\section{ITRU Cryptosystem Implementation}
We note that to fix $q$  one needs a bound for the largest possible value of the representation, so here if one only uses the letters from 'A' to 'Z' and 'a' to 'z', then the maximum is $122$. In the following SageMath implementation we will use $255.$ Moreover, we preform our implementation on the arbitrary message : Cryptanalysis.\\

\tcbset{ enhanced jigsaw}

\begin{tcolorbox}[colback=blue!5!white,colframe=blue!75!black, colbacktitle=blue!75!black,title=\texttt{ITRU Implementation Input}, breakable]
\mbox{\tiny{1}} \hspace{.3cm} $s = '$Cryptanalysis$'$ \\
\mbox{\tiny{2}} \hspace{.3cm} pretty\_ print($'$The message is:$ '$, s)\\
\mbox{\tiny{3}} \hspace{.3cm} $ r =8$\\
\mbox{\tiny{4}} \hspace{.3cm} $p =1000$\\
\mbox{\tiny{5}} \hspace{.3cm} $F=$ Set([$k$ for $k$ in range$(2,1000)$ if $\gcd(k,1000)==1$])\\
\mbox{\tiny{6}} \hspace{.3cm}  $f=F.$ random\_element()\\
\mbox{\tiny{7}} \hspace{.3cm} $S=$Set($[2..1000]$)\\
\mbox{\tiny{8}} \hspace{.3cm}  $g=S.$ random\_element()\\
\mbox{\tiny{9}} \hspace{.3cm}  $m=$[ord($k$) for $k$ in $s$]\\
\mbox{\tiny{10}} \hspace{.3cm} pretty\_print($'$ The ASCII code of the message :$'$, $m$)\\
\mbox{\tiny{11}} \hspace{.3cm} $q=$next\_prime($p*r*g+255*f$)\\
\mbox{\tiny{12}} \hspace{.3cm} $F_p=(1/f) \%  p$\\
\mbox{\tiny{13}} \hspace{.3cm} $Fq=(1/f)\% q$\\
\mbox{\tiny{14}} \hspace{.3cm} $ h=(p*Fq*g)\%q$\\
\mbox{\tiny{15}} \hspace{.3cm}  pretty\_print($ ' $ Large modulus :$', q$)\\
\mbox{\tiny{16}} \hspace{.3cm}  pretty\_print($' $ Public key :$',h$)\\
\mbox{\tiny{17}} \hspace{.3cm} pretty\_print($'$ Private key pair :$', (f,Fp)$)\\
\mbox{\tiny{18}} \hspace{.3cm} $e=[((r*h)+m[i])\%q$ for $i$ in $[0..len(m)-1]$]\\
\mbox{\tiny{19}} \hspace{.3cm}  pretty\_print($'$ The encrypted message :$',e$)\\
\mbox{\tiny{20}} \hspace{.3cm}  $a=[(f*e[i])\%q$ for $ i$ in $[0..len(e)-1]$]\\
\mbox{\tiny{21}} \hspace{.3cm} pretty\_print(html($r'\$f $\textbackslash cdot $e$ \textbackslash pmod$\{q\}\$ $ is : $ \$\% s \$'\% $latex($a$)))\\
\mbox{\tiny{22}} \hspace{.3cm} $C=[(F_p*a[l])\% p$ for $l$ in $[0..len(a)-1]$]\\
\mbox{\tiny{23}} \hspace{.3cm} pretty\_print(html($r'\$ F$\_p \textbackslash cdot $a$ \textbackslash pmod \{$q$\}\$  is : $\$ \% s \$ ' \%$ latex($C$)))\\
\mbox{\tiny{24}} \hspace{.3cm} $D=$[chr($k$) for $k$ in $C$]\\
\mbox{\tiny{25}} \hspace{.3cm} pretty\_print($'$ The original message :$', '$\hspace{.05cm}$'$.join($D$))
\end{tcolorbox}

\begin{tcolorbox}[colback=green!5!white,colframe=green!75!black, colbacktitle=green!75!black,title=\texttt{Output}, breakable]

The message is : Cryptanalysis\\
The ASCII code of the message :$ [67, 114, 121, 112, 116, 97, 110, 97, 108, 121, 115, 105, 115]$\\
Large modulus : $6186617$\\
Public key : $180058$\\
Private key pair :$ (73, 137)$\\
The encrypted message : $[1440531, 1440578, 1440585, 1440576, 1440580, 1440561,\\
1440574, 1440561, 1440572, 1440585, 1440579, 1440569, 1440579]$\\
$f \cdot e \pmod{q}$ is :$[6172891, 6176322, 6176833, 6176176, 6176468, 6175081, 6176030,$\\
$ 6175081, 6175884, 6176833, 6176395, 6175665, 6176395]$\\
$F_p \cdot a \pmod{p}$ is : $ [67, 114, 121, 112, 116, 97, 110, 97, 108, 121, 115, 105, 115]$\\
The original message : Cryptanalysis
\end{tcolorbox}

\section{ITRU Plaintext Recovery}

In this section we show how the ITRU cryptosystem can be attacked using a simple frequency analysis on the letters of cipher message. This attack is preformed with SageMath Software in which the plaintext is completely recovered only from the ciphertext and the public key with no need to have the private key. However, this attack technique can be applied on any encrypted message using the ITRU cryptosystem, let us preform this technique on the following paragraph from the article describing ITRU cryptosystem \cite{gaithuru2017itru} (without spaces):

\begin{tcolorbox}[enhanced, colback=gray!5!white,colframe=gray!75!black,fonttitle=\bfseries, breakable]
$'$\textbf{ThegoalofthisstudyistopresentavariantofNTRUwhichisbasedontheringof\\ integersasopposedtousingthepolynomialringwithintegercoefficients.We\\
showthatNTRUbasedontheringofintegers(ITRU),hasasimpleparameter\\
selectionalgorithm,invertibilityandsuccessfulmessagedecryption.We\\ describeaparameterselectionalgorithmandalsoprovideanimplementation\\
ofITRUusinganexample.ITRUisshowntohavesuccessfulmessagedecryption,\\
whichprovidesmoreassuranceofsecurityincomparisontoNTRU.}$'$
\end{tcolorbox}

If this paragraph is encrypted with the large modulus $q=1104427$ and the public key $h= 37619$, then the ciphertext starts as
$$
301036, 301056, 301053, 301055, 301063, 301049, 301060, 301063, 301054, . . . .
$$
In fact, there are $32$ different numbers appearing in the ciphertext these are between $300992$ and $301073.$
A simple frequency analysis with the function \texttt{frequency\_distribution()} provides the following data:

\begin{tcolorbox}[enhanced, colback=yellow!5!white,colframe=yellow!75!black,fonttitle=\bfseries, breakable]
\begin{align*}
[&(301056, 0.0380313199105145), (301057, 0.0850111856823266),\\
&(301060, 0.0313199105145414), (301061, 0.0290827740492170),\\
&(301062, 0.0648769574944072), (301063, 0.0738255033557047),\\
&(301064, 0.0313199105145414), (301066, 0.0536912751677852),\\
&(301067, 0.0850111856823266), (301068, 0.0693512304250559),\\
&(301069, 0.0201342281879195), (301070, 0.0111856823266219),\\
&(301071, 0.0111856823266219), (301072, 0.00223713646532438),\\
&(301073, 0.0134228187919463), (300992, 0.00223713646532438),\\
&(300993, 0.00223713646532438), (300996, 0.00671140939597315),\\
&(300998, 0.00894854586129754), (301025, 0.00671140939597315),\\
&(301030, 0.00671140939597315), (301034, 0.0134228187919463),\\
&(301036, 0.0156599552572707), (301037, 0.0134228187919463),\\
&(301039, 0.00447427293064877), (301049, 0.0693512304250559),\\
&(301050, 0.00894854586129754), (301051, 0.0357941834451902),\\
&(301052, 0.0246085011185682), (301053, 0.109619686800895),\\
&(301054, 0.0223713646532438), (301055, 0.0290827740492170)]
\end{align*}
\end{tcolorbox}

We see that the number $301053$ appears the most in the ciphertext. Therefore, $301053$
represents either $’e’, ’a’$ or $’t’$. If it is $’e’$, then we apply the formula 
$$
c_i-300952,
$$
where $c_i$ represents the ciphertext blocks in the ASCII character code for all $i$. Thus, we get a sequence of numbers starting with
$$
84, 104, 101, 103, 111, 97, 108, 111, 102,....
$$
Finally,  if we consider it as a sequence of ASCII codes and determine the corresponding plaintext,
then we get the encoded message.
\section{Acknowledgments}
 The research was supported in part by grants K115479 and K128088 (Sz.T.) of the Hungarian National Foundation for Scientific Research. The work of H.~R. Hashim was supported by the Stipendium Hungaricum Scholarship.

\bibliography{all}

\begin{thebibliography}{10}

\bibitem{zbMATH01955782}
William~D. {Banks} and Igor~E. {Shparlinski}.
\newblock {A variant of NTRU with non-invertible polynomials.}
\newblock In {\em {Progress in cryptology -- INDOCRYPT 2002. Third
  international conference on cryptology in India, Hyderabad, India, December
  16--18, 2002. Proceedings}}, pages 62--70. Berlin: Springer, 2002.

\bibitem{camara2018dtru1}
M.~G. Camara, De. Sow, and Dj. Sow.
\newblock Dtru1: First generalization of ntru using dual integers.
\newblock {\em International Journal of Algebra}, 12(7):257--271, 2018.

\bibitem{zbMATH05250294}
Michael {Coglianese} and Bok-Min {Goi}.
\newblock {MaTRU: A new NTRU-based cryptosystem.}
\newblock In {\em {Progress in cryptology -- INDOCRYPT 2005. 6th international
  conference on cryptology in India, Bangalore, India, December 10--12, 2005,
  Proceedings}}, pages 232--243. Berlin: Springer, 2005.

\bibitem{10.1007/3-540-6}
D.~Coppersmith and A.~Shamir.
\newblock Lattice attacks on ntru.
\newblock In Walter Fumy, editor, {\em Advances in Cryptology --- EUROCRYPT
  '97}, pages 52--61, Berlin, Heidelberg, 1997. Springer Berlin Heidelberg.

\bibitem{MR798552}
T.~ElGamal.
\newblock A public key cryptosystem and a signature scheme based on discrete
  logarithms.
\newblock {\em IEEE Trans. Inform. Theory}, 31(4):469--472, 1985.

\bibitem{gaborit:inria}
P.~Gaborit, J.~Ohler, and P.~Sol{\'e}.
\newblock {CTRU, a polynomial analogue of NTRU}.
\newblock Technical Report RR-4621, {INRIA}, November 2002.

\bibitem{gaithuru2017itru}
J.~N. Gaithuru, M.~Salleh, and I.~Mohamad.
\newblock Itru: Ntru-based cryptosystem using ring of integers.
\newblock {\em International Journal of Innovative Computing}, 7(1), 2017.

\bibitem{zbMATH01722678}
C.~{Gentry}.
\newblock {Key recovery and message attacks on NTRU-composite.}
\newblock In {\em {Advances in cryptology - EUROCRYPT 2001. 20th international
  conference on theory and application of cryptographic techniques, Innsbruck,
  Austria, May 6--10, 2001. Proceedings}}, pages 182--194. Berlin: Springer,
  2001.

\bibitem{zbMATH06590793}
Henri {Gilbert}, Dipankar {Gupta}, Andrew {Odlyzko}, and Jean-Jacques
  {Quisquater}.
\newblock {Attacks on Shamir's `RSA for paranoids'.}
\newblock {\em {Inf. Process. Lett.}}, 68(4):197--199, 1998.

\bibitem{MR1630399}
O.~Goldreich, S.~Goldwasser, and S.~Halevi.
\newblock Public-key cryptosystems from lattice reduction problems.
\newblock In {\em Advances in cryptology---{CRYPTO} '97 ({S}anta {B}arbara,
  {CA}, 1997)}, volume 1294 of {\em Lecture Notes in Comput. Sci.}, pages
  112--131. Springer, Berlin, 1997.

\bibitem{zbMATH01186948}
J.~{Hoffstein}, J.~{Pipher}, and J.~H. {Silverman}.
\newblock {NTRU: A ring-based public key cryptosystem}.
\newblock In {\em {Algorithmic number theory. 3rd international symposium,
  ANTS-III, Portland, OR, USA, June 21--25, 1998. Proceedings}}, pages
  267--288. Berlin: Springer, 1998.

\bibitem{10.1007/978-3-540-74143-5_9}
Nick Howgrave-Graham.
\newblock A hybrid lattice-reduction and meet-in-the-middle attack against
  ntru.
\newblock In Alfred Menezes, editor, {\em Advances in Cryptology - CRYPTO
  2007}, pages 150--169, Berlin, Heidelberg, 2007. Springer Berlin Heidelberg.

\bibitem{10.1007/3-54}
{\'E}liane Jaulmes and Antoine Joux.
\newblock A chosen-ciphertext attack against ntru.
\newblock In Mihir Bellare, editor, {\em Advances in Cryptology --- CRYPTO
  2000}, pages 20--35, Berlin, Heidelberg, 2000. Springer Berlin Heidelberg.

\bibitem{joye1997importance}
Marc Joye and Jean-Jacques Quisquater.
\newblock On the importance of securing your bins: The
  garbage-man-in-the-middle attack.
\newblock In {\em Proceedings of the 4th ACM conference on Computer and
  communications security}, pages 135--141, 1997.

\bibitem{6329211}
A.~A. {Kamal} and A.~M. {Youssef}.
\newblock A scan-based side channel attack on the ntruencrypt cryptosystem.
\newblock In {\em 2012 Seventh International Conference on Availability,
  Reliability and Security}, pages 402--409, 2012.

\bibitem{eprint-2015-26958}
A.~H. Karbasi and R.~E. Atani.
\newblock Iltru: An ntru-like public key cryptosystem over ideal lattices.
\newblock {\em IACR Cryptology ePrint Archive}, 2015:549, 2015.

\bibitem{articlepairtru}
A.~H. Karbasi, R.~E. Atani, and S.~E. Atani.
\newblock Pairtru: Pairwise non-commutative extension of the ntru public key
  cryptosystem.
\newblock {\em International Journal of Information Security Science}, 8:1--10,
  03 2018.

\bibitem{MR866109}
Neal Koblitz.
\newblock Elliptic curve cryptosystems.
\newblock {\em Math. Comp.}, 48(177):203--209, 1987.

\bibitem{kouzmenko2006}
R~Kouzmenko.
\newblock Generalizations of the ntru cryptosystem.
\newblock {\em Diploma Project, {\'E}cole Polytechnique F{\'e}d{\'e}rale de
  Lausanne,(2005--2006)}, 2006.

\bibitem{MR3989003}
Zhen Liu, Yanbin Pan, and Zhenfei Zhang.
\newblock Cryptanalysis of an {NTRU}-based proxy encryption scheme from
  {ASIACCS}'15.
\newblock In {\em Post-quantum cryptography}, volume 11505 of {\em Lecture
  Notes in Comput. Sci.}, pages 153--166. Springer, Cham, 2019.

\bibitem{malekian2011qtru}
E.~Malekian, A.~Zakerolhosseini, and A.~Mashatan.
\newblock Qtru: Quaternionic version of the ntru public-key cryptosystems.
\newblock {\em ISeCure}, 3(1), 2011.

\bibitem{1978DSNPR}
R.~J. {McEliece}.
\newblock {A Public-Key Cryptosystem Based On Algebraic Coding Theory}.
\newblock {\em Deep Space Network Progress Report}, 44:114--116, January 1978.

\bibitem{zbMATH05021034}
Tommi {Meskanen} and Ari {Renvall}.
\newblock {A wrap error attack against NTRUEncrypt.}
\newblock {\em {Discrete Appl. Math.}}, 154(2):382--391, 2006.

\bibitem{Micciancio2005}
D.~Micciancio.
\newblock {\em Closest Vector Problem}, pages 79--80.
\newblock Springer US, Boston, MA, 2005.

\bibitem{zbMATH05242559}
Petros {Mol} and Moti {Yung}.
\newblock {Recovering NTRU secret key from inversion oracles.}
\newblock In {\em {Public key cryptography -- PKC 2008. 11th international
  workshop on practice and theory in public-key cryptography, Barcelona, Spain,
  March 9--12, 2008. Proceedings}}, pages 18--36. Berlin: Springer, 2008.

\bibitem{zbMATH05723624}
M.~{Nevins}, C.~{KarimianPour}, and A.~{Miri}.
\newblock {NTRU over rings beyond \(\mathbb{Z}\).}
\newblock {\em {Des. Codes Cryptography}}, 56(1):65--78, 2010.

\bibitem{10.1145}
David Nu\~{n}ez, Isaac Agudo, and Javier Lopez.
\newblock Ntrureencrypt: An efficient proxy re-encryption scheme based on ntru.
\newblock In {\em Proceedings of the 10th ACM Symposium on Information,
  Computer and Communications Security}, ASIA CCS ’15, page 179–189, New
  York, NY, USA, 2015. Association for Computing Machinery.

\bibitem{standard1977l}
National~Bureau of~Standards.
\newblock Data encryption standard.
\newblock {\em FIPS Publication 46, U.S. Department of Commerce}, 1977.

\bibitem{advanced2001}
National~Institute of~Standards and Technology.
\newblock Advanced encryption standard.
\newblock {\em FIPS Publication 197, U.S. Department of Commerce}, 2001.

\bibitem{10.1007/978}
Y.~Pan and Y.~Deng.
\newblock A general ntru-like framework for constructing lattice-based
  public-key cryptosystems.
\newblock In Souhwan Jung and Moti Yung, editors, {\em Information Security
  Applications}, pages 109--120, Berlin, Heidelberg, 2012. Springer Berlin
  Heidelberg.

\bibitem{eprint-2003-11720}
John Proos.
\newblock Imperfect decryption and an attack on the ntru encryption scheme,
  2003.
\newblock japroos@math.uwaterloo.ca 12059 received 7 Jan 2003.

\bibitem{MR700103}
R.~L. Rivest, A.~Shamir, and L.~Adleman.
\newblock A method for obtaining digital signatures and public-key
  cryptosystems.
\newblock {\em Comm. ACM}, 21(2):120--126, 1978.

\bibitem{key:article}
Gurpreet S. and Supriya.
\newblock A study of encryption algorithms (rsa, des, 3des and aes) for
  information security.
\newblock {\em International Journal of Computer Applications}, 67(19):33--38,
  2013.

\bibitem{zbMATH02091528}
Tanya~E. {Seidel}, Daniel {Socek}, and Michal {Sramka}.
\newblock {Parallel symmetric attack on NTRU using non-deterministic lattice
  reduction.}
\newblock {\em {Des. Codes Cryptography}}, 32(1-3):369--379, 2004.

\bibitem{5341656}
X.~{Shen}, Z.~{Du}, and R.~{Chen}.
\newblock Research on ntru algorithm for mobile java security.
\newblock In {\em 2009 International Conference on Scalable Computing and
  Communications; Eighth International Conference on Embedded Computing}, pages
  366--369, Sep. 2009.

\bibitem{doi:10.1002/sec.1693}
S.~Singh and S.~Padhye.
\newblock Generalisations of ntru cryptosystem.
\newblock {\em Security and Communication Networks}, 9(18):6315--6334, 2016.

\bibitem{sage}
W.~A. Stein et~al.
\newblock {\em {S}age {M}athematics {S}oftware ({V}ersion 9.0)}.
\newblock The Sage Development Team, 2020.
\newblock {\tt http://www.sagemath.org}.

\bibitem{MR2221458}
J.~Talbot and D.~Welsh.
\newblock {\em Complexity and cryptography}.
\newblock Cambridge University Press, Cambridge, 2006.
\newblock An introduction.

\bibitem{articlebitru}
H.~Yassein and N.~Al-Saidi.
\newblock Bitru: Binary version of the ntru public key cryptosystem via binary
  algebra.
\newblock {\em International Journal of Advanced Computer Science and
  Applications}, 12 2016.

\end{thebibliography}

\bigskip
\hrule
\bigskip

\noindent 2010 {\it Mathematics Subject Classification}:
Primary 11D25; Secondary 11B37, 11B39, 11A63, 11J86.

\noindent \emph{Keywords: }
Lucas sequences, Diophantine equations, Elliptic curves, Repdigits.

\bigskip
\hrule
\bigskip

\end{document}